\documentclass[proof]{WileyASNA-v1}

\articletype{Article Type}%

\received{13 October 2020}
\revised{18 October 2020}
\accepted{22 October 2020}

\begin{document}

\title{Neutron Star {\it versus} Neutral Star:\\
{\it On the 90$^{th}$ anniversary of Landau's publication in astrophysics}}
\author[aff1,aff2]{Renxin Xu*}

\address[aff1]{School of Physics and State Key Laboratory of Nuclear Physics and Technology, Peking University, Beijing 100871, China}
\address[aff2]{Kavli Institute for Astronomy and Astrophysics, Peking University, Beijing 100871, China}

\corres{*Renxin Xu, School of Physics, Peking University, Beijing 100871, China. \\
\email{r.x.xu@pku.edu.cn}}

\abstract{%
In the late age of developing quantum mechanics, Lev Landau, one of the distinguished players, made great efforts to understand the nature of matter, even stellar matter, by applying the quantum theory.
Ninety years ago, he published his idea of ``neutron'' star, which burst upon him during his visit over Europe in the previous year.
The key point that motivated Landau to write the paper is to make a state with lower energy for ``gigantic nucleus'', avoiding extremely high kinematic energy of electron gas due to the new Fermi-Dirac statistics focused hotly on at that time.
Landau had no alternative but to neutronize/neutralize by ``combining a proton and an electron'', as electron and proton were supposed to be elementary before the discovery of neutron.
However, our understanding of the Nature has fundamentally improved today, and another way (i.e., strangeonization) could also embody neutralization and thus a low-energy state that Landau had in mind, which could further make unprecedented opportunities in this multi-messenger era of astronomy.
Strangeon matter in ``old'' physics may impact dramatically on today’s physics, from compact stars initiated by Landau, to cosmic rays and dark matter.
In this essay, we are making briefly the origin and development of neutron star concept to reform radically, to remember Landau's substantial contribution in astrophysics and to recall those peculiar memories.
}%
\keywords{pulsar, neutron star, dense matter, elementary particles}

\maketitle

\section{Introduction}
The Mandela effect\footnote{
Interestingly, a similar effect has already been included in a Chinese idiom, ``to circulate erroneous reports by someone wrongly informed'' (yǐ é chuán é) in as early as the southern Song dynasty, to be very popular in the social contacts nowadays.
} %
is miraculous in psychology and media studies, which could be the result of
wrong combination of memory fragments under memory reconstruction.
Unfortunately, it affects also tremendously the historical memory of the neutron star concept, as will be explained briefly here.
{\it Researchers} are encouraged to think seriously and independently to read the  literatures, drawing a lesson from the mix-up memory, especially during this information age.

Yes, it should be recommended for one to focus on real things rather than only fashions in the mainstream, since the latter could be strongly coupled with the media transportation.
Nonetheless, there is some fundamental differences between natural sciences and social actions: the laws of nature are objective, never minding the social emotions!
In this sense, the publication of Landau's idea about neutron stars could also be an excellent example for {\it educational} meaning.

Let's make an effort in this aspect.
In order to clarify the fact and its implications, the sciences relevant to the publication by~\citet{Landau1932} on stars, including the past and future, would be reformed from the bottom in this contribution, on the 90$^{\rm th}$ anniversary of Landau's presenting his original idea.

\section{The magnificent era: physics from macroscopic to microscopic, and eventually to the quantum}

\begin{figure}[h!]
\centering
\includegraphics[scale=0.8]{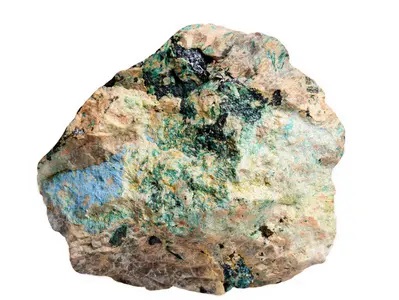}
\caption{A rock containing copper (Malachite). Normal atom matter at pressure free is condensed by the electromagnetic (or simply electric) force, while the strangeon matter, to be explained in \S6, is by the fundamental strong interaction. Multiscale forms can exist for both kinds of condensed matter, the electric and the strong ones.
} %
\label{fig:Cu}
\end{figure}

The developments of social civilization are driven by the curiosity of human beings.
What's the nature of our material world (e.g., the rock shown in Fig.~\ref{fig:Cu})? Why does it exist?
The basic unit of normal matter was speculated even in the pre-Socratic period of the Ancient era (the basic stuff was
hypothesized to be indestructible\footnote{
Oppositely,  substances in our daily life are thought to be destructible in ancient China, as was recorded in Tianxia Pian by Zhuang Zi ($\sim$ 369BC$-$286BC): `` {\it Taking a stick of wood with finite extent, one cuts half each day;
it is expected to last for an infinitely long time}''.
} %
``atoms'' by Democritus, $\sim$ 460BC$-$370BC), but it was a belief that symmetry, which is well-defined in
mathematics, should play a key role in understanding the material structure, such as the Platonic solids (i.e., the five
regular convex polyhedrons).

The philosophical thinking above became gradually related to practical sciences of understanding the thermal phenomenon more than two thousand years later ($\sim 1700$AD$-1900$AD), resulting in the first industrial revolution.
Thermodynamics focuses on the thermal properties of matter from a macro perspective, while statistical physics studies the global behavior of a huge number of particles in molecular models. %
Both approaches reach the same end from different directions.
Many forerunners offered wisdom for understanding atom matter (e.g., Fig.~\ref{fig:Cu}), especially the realist among the main figures, Ludwig Edward Boltzmann ($1844-1906$), who ``stands as a link between two other great theoretical physicists: James Clerk Maxwell in the 19$^{\rm th}$ century and Albert Einstein in the 20$^{\rm th}$''~\citep{Boltzmann}.
This is a pretty big step to explore the fundamental units of Nature, leading us into a scientific world from atoms to subatoms.

Ahh...What's the nature of atom?
In fact, atoms are {\it destructible} since J. J.Thomson addressed that the so-called ``cathode rays'' leaving from atoms are actually negatively charged  but extremely light units (i.e., electrons) in 1897.
The other part of atom is believed to be positive, massive and small nucleus after Ernest Rutherford explained the large angle scattering of the Geiger–Marsden experiment in 1911.
However, such a ``nucleus+electron(s)'' atom cannot  be stable at all in classical electrodynamics, though Niels Bohr assumed that an atom should not collapse by introducing a quantum angular momentum (ie. Bohr's distinct energy states) in 1913, that calls for a new era of completely different mechanics.

It is now well known that the revolution of quantum mechanics unveiled the facts of atom.
Beginning with the study of black-body radiation, especially the Planck's formulae of spectrum in 1900 (it is worth noting that Boltzmann's ideas were central to Planck's analysis), we know that light is particle-like, each one with energy (Einstein in 1905) and momentum (Compton in 1923), though it is wave-like in classical electromagnetism.
This peculiar wave-particle duality of light was extended for all the other ``particles'', particularly the electron, by de Broglie in 1924.
Eventually, {\it quantum mechanics} (the
matrix mechanics by Heisenberg in 1925 and the wave mechanics by Schr\"odinger in 1926) was established, which reproduces quantitatively the eigen energies of the ground state and the excited ones, providing essentially a solid foundation for Bohr’s distinct energy states.

However, why don't all the electrons in a stable atom go to the ground quantum state?
Physicists cannot understand until a ``new'' statistics (i.e., the third one) was discovered.

\section{The ``New'' Statistics and its implications}

By the end of the 19$^{\rm th}$ century and at the beginning of the 20$^{\rm th}$, Maxwell-Boltzmann (MB) statistics, successful in explaining the thermodynamics of gas and sometimes of liquid and solid, was believed without doubt until Einstein received a letter from Bose in 1924, about revisiting the puzzling black-body radiation.
Today, we know that Bose-Einstein (BE) statistics is for a quantum system of identical particles (i.e., Bosons) whose states are exchangeable.
Certainly, the thermal system of photons is a typical example.

The third statistics was established soon after Pauli proposed his exclusion principle for electrons in 1925, explaining the periodic law of chemical elements.
This statistical property of electron, differing from that of photon, was noted\footnote{
It is also said that Pascual Jordan  (1902-1980) was the first one who proposed the new statistics in 1925~\citep{Schucking1999}. This sad story happened due to Max Born's negligence.
Jordan referred to his own brainchild as ``the Pauli statistics''.
} %
independently by Fermi and Dirac in 1926, marking the birth of the new statistics, i.e., the Fermi-Dirac (FD) one.

FD statistics is powerful in understanding the Nature, and the first three examples are listed as following.

1, To solve a problem in the ``Heaven'' by~\cite{1926MNRAS..87..114F}: What's nature of white dwarfs with extremely high compactness compared to normal main-sequence stars?
In FD statistics, cold electron gas in a white dwarf, even at zero temperature, could still contribute significant pressure, while the pressure in main-sequence stars is only  of MB statistics.
In fact, at low temperature limit, the pressure of FD statistics is surprisingly larger than that of traditional MB statistics, because the kinematic energy of dense electrons is remarkably high, as illustrated by the Heisenberg relation $\Delta x\cdot \Delta p\sim \hbar$, with $\Delta x$ the separation between electrons, $\Delta p$ the order of momentum and $\hbar$ the Planck constant.

2, To solve a problem on the ``Earth'' by~\cite{Sommerfeld1927}: Why is the specific heat of electrons in metal negligible?
Although the Drude-Lorentz model (1900-1909) based on MB statistics may explain the electrical and
thermal conductivities of metal, only a small part of electrons near the Fermi surface, $\sim kT/E_{\rm f}$, would contribute to the heat capacity in FD statistics, with $k$ the Boltzmann constant, $T$ the temperature and $E_{\rm f}$ the Fermi energy of electron which is much larger than $kT$ at room temperature.
The reason for this is also that the kinematic energy of an electron in comparably cold and dense matter is remarkably high.
This was testing the capability of quantum theory on problems beyond atoms, but the point of extremely high energy in cold and dense electron gas is meaningful again!

3, To make an effort of understanding the magnetism, to be so familiar a phenomenon but also a test of FD statistics.
The early history of quantum theory of solids, especially of metals was well summarized by~\cite{Hoddeson1987}.
\cite{Pauli1927}, for the first time, attempted to explain paramagnetism with FD statistics, but the others experimentally observed (e.g., diamagnetism, magnetoresistance, the
Hall effect)  were still puzzling.
As one of the visiting fellows, Landau joined in Pauli's group in 1929, and away from the group, he published his solution of the full quantum-mechanical problem of electrons orbiting in a magnetic field, explaining diamagnetism~\citep{Landau1930}.
This research was related to Landau's trip to Europe, as will be discussed in the next section.

\section{Lev Landau and his scientific trip over Europe}

The west side of Europe is a vital source of modern sciences, which was certainly attracting young students with a strong spirit of inquiry about the Nature.
Although the international exchanges become more and more convenient, there are still two major ways of reaching the center of science even today.
Two archetypes of the successful Asian-born youngsters of genius, during the later period of developing quantum theory in 1930s, are presented as following.

One is for Subrahmanyan Chandrasekhar (1910$-$1995), born in Indian~\citep{2011ffbh.book.....S} as a part within the British Commonwealth (about six thousand kilometers away from the center of science).
As one of the undergraduates at Presidency college, Chandra listened to the speech lectured by Sommerfeld in the autumn of 1928 and discussed afterward about FD statistics, leading him to publish the first and undergraduate research~\citep{1929RSPSA.125..231C} on Compton scattering of moving electrons which obey FD statistics (communicated by R. H. Fowler), to supplement Dirac's work on Compton scattering of moving electrons with Maxwellian distribution in hot stellar atmosphere~\citep{1925MNRAS..85..825D}.

On his two-and-a-half week journey from India to Cambridge in July 1930, Chandra  recognized that his supervisor's calculation~\citep{1926MNRAS..87..114F} was based on non-relativistic energy momentum relation, but the degenerate electrons could be relativistic in massive white dwarfs.
Chandra submitted this research titled ``The maximum mass of ideal white dwarfs'' ($M_{\rm max}=0.91M_\odot$, applying the theory of polytropic gas spheres) to ApJ on 12$^{\rm nd}$ November 1930, published in the following July~\citep{1931ApJ....74...81C}, known as the ``Chandrasekhar limit'' in text books.
Chandra was very productive in astrophysics, from hydrodynamics, magnetohydrodynamics, to general relativity and black hole.

The other is for Lev Landau (1908$-$1968), born in Baku of Azerbaijan~\citep{Lifshitz1969}, one of the former Soviet Union republics (about four thousand kilometers away from the center of science).
Landau started to travel abroad in October 1929, initially supported by the People's Commissariat of Education, worked in Denmark, Great Britain and Switzerland, and returned to Leningrad Physicotechnical Institute in March 1931.
Besides the theory of the diamagnetism of an electron gas, Landau was thinking the relativistic electrons in white dwarfs during his third times of visiting Bohr's group in Copenhagen when his Soviet stipend had run out but Bohr helped obtain a Rockefeller Fellowship which allowed Landau to prolong his stay.
During this time from February to March in 1931, Landau, Bohr and Rosenfeld discussed a paper written by Landau but not published, nonetheless Landau submitted the manuscript in January 1932 and published in February~\citep{Landau1932}, the same month when the discovery of the neutron was announced~\citep{2013PhyU...56..289Y}.

The key point of~\cite{Landau1932} is also relevant to the highly kinematic energy of dense electron gas, as sampled in \S3 on FD statistics.
Can any physical mechanism cancel the kinematic energy of electrons (typically $\hbar c/\Delta x\sim 300$ MeV if electrons keeps there) to form a stable state with lower energy?
Landau provided a way to do this, by combining ``protons and electrons in atomic nuclei very close together'', and he ``expect that this must occur when the density of matter becomes so great that atomic nuclei come in close contact, forming one gigantic nucleus''.
However, he didn't define {\it gigantic nucleus}: where is the boundary of baryon number, $A_{\rm c}$, between normal microscopic and gigantic nuclei?
Certainly a nucleus with stellar mass (baryon number $A\sim 10^{57}$) is gigantic since the corresponding number $A<300$ for atomic nuclei of chemical elements, but the critical number $A_{\rm c}$ could be somewhere between numbers differing $\sim 55$ orders of magnitude!

The ``neutron star'' concept superficially anticipated by Landau is not surprising, though before the discovery of neutron.
Before the establishment of quantum mechanics, it is a general idea that neutral doublet (combination of an electron and a proton, thus to differentiate cementing electrons from planetary ones in an atom) may exist in an atomic nucleus~\citep[e.g.,][]{Rutherford1920,Harkins1920},  postulating``novel properties'': to move freely through matter and to be difficult to detect.
This specific feature of weak interaction resembles that of dark matter with indirect evidence by gravity.
\cite{Harkins1921} renamed doublet ``neutron'', which was discovered by~\cite{Chadwick1932}.
Today, neutron imaging are very useful in material science and engineering, as well as biology~\citep{Anderson2009}.

Even before the work by~\cite{1931ApJ....74...81C}, Edmund C. Stoner (1899$-$1968), known for his work on the origin and nature of magnetism, had also noted a limiting mass of white dwarfs ($M_{\rm max}\sim M_\odot$) in the uniform density approximation~\citep{Stoner1930}, considering that the total energy of kinetic electrons and stellar gravity should be minimum.
\cite{Landau1932} recalculated the mass limit of white dwarfs ($M_{\rm max}= 1.5 M_\odot$) too, which could be expressed in terms
of fundamental constants.
Let’s consider a self-gravitational star with mass $M$ and radius $R$, consisting of basic unit with mass $m_0$ and total number $N$.
With the overall energy, $E\simeq N\cdot (\hbar cN^{1/3}-Gm_0^2 N)/R$, one has then the mass limit $M_{\rm max}\sim N_{\rm max}\cdot m_0 \simeq M_{\rm p}^3/m_0^2$, with the Planck mass $M_{\rm p}=\sqrt{\hbar c/G}=1.2\times 10^{19}$ GeV, otherwise $R=0$ if $N>N_{\rm max}$.
Estimating the scale of each unit to be the Compton wavelength $\lambda_0 \sim \hbar/(m_0c)$, we have thus the typical radius $R\simeq \lambda_0\cdot N_{\rm max}^{1/3}$.
For $m_0\sim 1$ GeV, we have characteristic magnitudes
for compact stars: $N_{\rm max} \sim 10^{57}$ and $M_{\rm max}\sim M_\odot$.
It is evident that both quantum physics and special/general relativity should participant in this study, as was indicated by the physical constants of $\hbar$, $c$ and $G$.

Landau's idea develops then, especially after the discovery of pulsars, and becomes very elaborate models of normal neutron stars in the mainstream.
But there is still a fundamental misunderstanding about the origin of the idea.
Rosenfeld told an anecdote in 1974 that Landau improvised the concept of neutron stars in a discussion with Bohr and Rosenfeld just after the news
of the discovery of the neutron reached Copenhagen in February 1932, which is noted in the famous text book by~\cite{1983bhwd.book.....S}.
This false memory of Rosenfeld, presented in a talk 43 years later, matters seriously in the academic society, that is another example caused by the Mandela effect.

\section{Landau's Faith: Neutron Star or Neutral Star?}

As discussed above, Landau's mind was actually to make a neutral star when density becomes so large that atomic nuclei come in close contact.
It is worth noting, however, that the mass of Landau's gigantic nucleus~\citep{Landau1932} should not always be $>M_\odot$, but probably $\ll M_\odot$, i.e., maybe $A_{\rm c}\ll 10^{57}$.
Landau noticed that a white dwarf with mass $>M_{\rm max}$ cannot be stable against collapsing to a gigantic nucleus, but low-mass such a nucleus,  surrounded by normal ions and electrons, could still be possible {\it if} the laws of quantum mechanics (and thus FD statistics) are violated.
Many authors did then follow along the line of Landau (1932,1938), to model neutron stars tested by astronomical observations, setting chemical equilibrium between phases in a star, since afterwards Landau published a special paper focused again on this topic, trying to solve the problem of stellar energy~\citep{Landau1938}.

We have to say that only neutronization can achieve Landau's goal of neutralization during the era when proton and neutron are supposed to be fundamental particles.
But a dangerous law, unknown at Landau's time, lurks in this option: the nuclear {\it symmetry energy}!
Although neutronization can remove energetic electrons, an extremely high asymmetry of the resultant isospin will contribute significant symmetry energy.
Meanwhile, due to the large asymmetry of isospin, dense electron gas would be necessary to suppress the $\beta$-decay of neutron to proton, that requires normal matter of stellar crust to meet the standard of such an electron density.

\begin{figure*}[h!]
\centering
\includegraphics[scale=0.5]{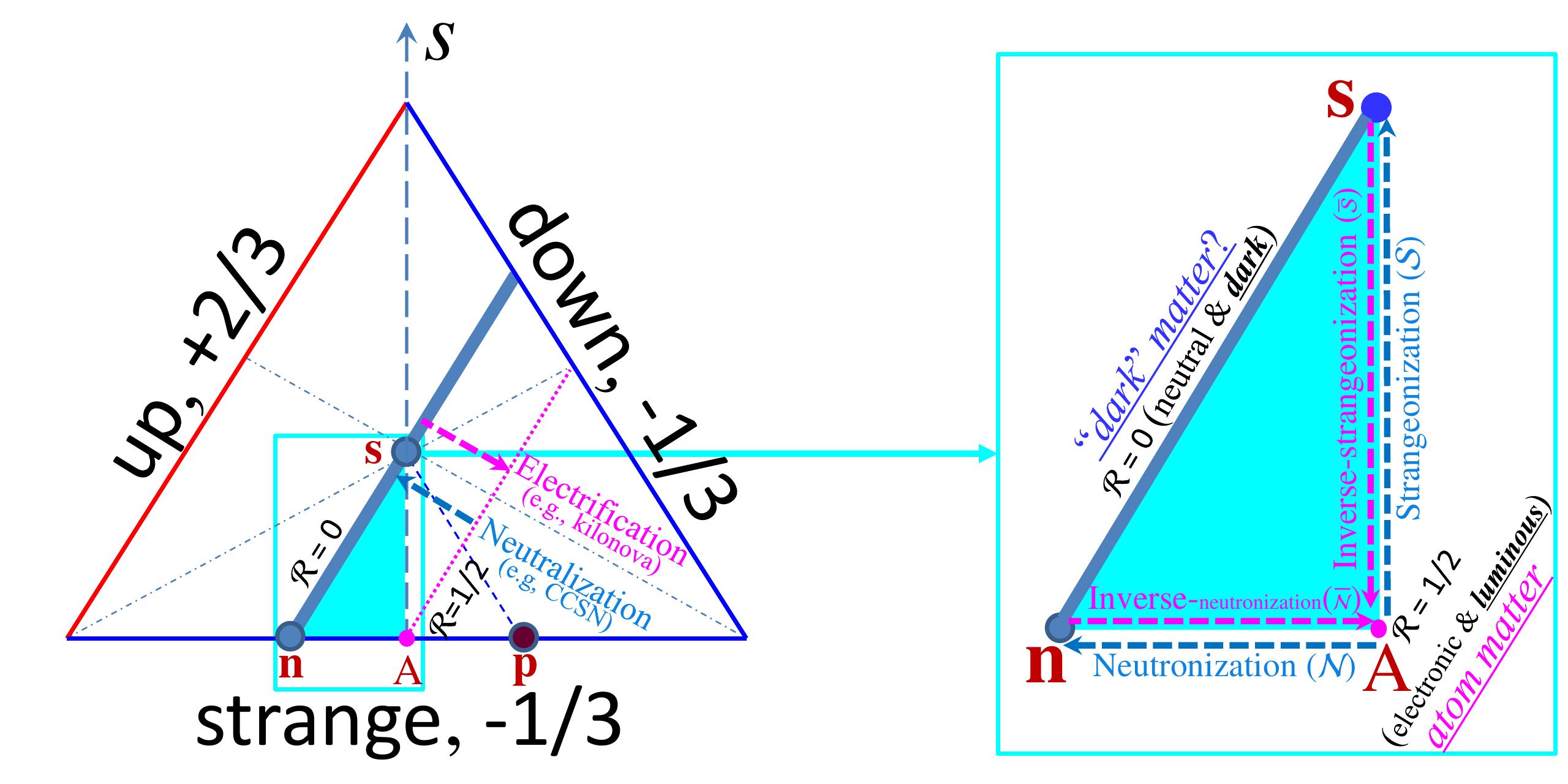}
\caption{An updated triangle of light-quark flavours~\citep{XuRX2020}. The points inside the triangle define the states with certain quark number densities of three flavours (\{$n_{\rm u}, n_{\rm d}, n_{\rm s}$\} for up, down and strange quarks), measured by the heights of one point to one of the triangle edges.
Axis $S$ denotes strangeness, with perfect isospin symmetry.
Normal nuclei are around point ``A'', conventional neutron stars are in point ``n'', extremely unstable proton stars are in point ``p'', and strange stars (both strange quark star and strangeon star) in point ``s'' in the center of the triangle with $n_{\rm u}=n_{\rm d}=n_{\rm s}$.
The charge-mass-ratio of quarks at point ``A'' is $\mathcal{R}\simeq 1/2$, but $\mathcal{R}=0$ on the line of ``ns''; the former is ``electronic'' and ``luminous'', while the latter could hardly be detected because of neutrality and would take a ``dark'' role as matter.
The zoomed $\triangle$snA shown in the right indicates that, supernovae explosions (from point ``A'' to a point in line ``ns'') realize neutronization ($\mathcal{N}$) or strangeonization ($\mathcal{S}$), and kilonovae realize inverse-neutronization ($\bar{\mathcal{N}}$, from point ``n'' back to point ``A'') or inverse-strangeonization ($\bar{\mathcal{S}}$, ``s$\rightarrow$A'').
} %
\label{fig:triangle}
\end{figure*}

Should the neutronization be the only way to neutralize in a gigantic nucleus?
Certainly, alternative way to neutralize is possible in the standard model of particle physics.
Due to the strong coupling of $\alpha_{\rm s}\sim 1$, quarks are bound inside a system (e.g., a nucleon) with scale of $\le 1$ fm, and the typical energy is then $E_{\rm scale}\sim \hbar c/\Delta x\simeq 0.5$ GeV, with $\Delta x$ the separation between quarks.
This means that, in the first step, we may ignore heavy flavours of quarks (c,t,b), and could take advantage of a triangle diagram (Fig.~\ref{fig:triangle}) for the light  flavors of quarks (u,d,s).
Normal atomic nuclei are around point ``A''.
Landau's way ends at point ``n'', while the other way could be from ``A'' to ``s''.
Both ways above are of neutralization, in response to the original intention of~\cite{Landau1932}.
At the quark level, the net results of neutronization and strangeonization are $u + e\rightarrow d + \nu_e$, $u + e\rightarrow s + \nu_e$, respectively.

Light quark-flavour symmetry is restored at point ``s'', but the building units could be either quarks~\cite[e.g.,][]{Witten1984} or strangeons~\citep{Xu2003,LX2017}.
The former is based on the pertubative strong interaction between quarks, while the latter on the non-pertubative.
It is well-known that the nature of an atomic nucleus, as microscopic strong matter, is determined by non-pertubative QCD (quantum chromodynamics), and one may simply argue that a gigantic nucleus at pressure free could also be the similar.
In this sense, a strangeon is an analogy of a nucleon, but only changing the number of valence-quark flavours from 2 to 3, i.e., {\bf 2}-flavoured nucleon while {\bf 3}-flavoured strangeon.

In a word, atomic nuclei are {\bf 2}-flavoured, but a gigantic nucleus could be {\bf 3}-flavoured. Where is the boundary?
One should be embarrassed for answering the question about $A_{\rm c}$, which is essentially determined by both the weak and the strong interactions, especially the latter in non-perturbative regime.
Frankly speaking, one should ask Landau for $A_{\rm c}$ since he proposed ``gigantic nucleus'' 90 years ago, if he could still be healthy today and if one does not care about Landau's possible feeling of awkwardness.
Nonetheless, from the weak side, an electron becomes relativistic if localized in the Compton wavelength $\hbar/(m_{\rm e}c)\sim 10^3$ fm, one may have $A_{\rm c }\sim 10^9$.
From the strong side, one could have $A_{\rm c } \ge 10^2$ in a liquid drop model.
We may then estimate a value of $A_{\rm c }$ between $10^3$ and $10^{10}$.

\section{Neutron {\it versus} Strangeon}

Although both neutron and strangeon are neutral, two differences between them are worth noting.

1, {\it Flavour Symmetry}.
Adopting a phenomenological approach,
nuclear physicists introduce a symmetry energy representing
the symmetry between the proton and neutron of a stable
nucleus, which is essentially the balance of two flavors of
quarks (u,d), but the underlying physics is yet to be well
understood.
For {\bf 2}-flavoured nuclei, it's impossible to satisfy both the neutralization and the flavour-symmetry, but simply a straightforward task for {\bf 3}-flavoured nuclei.~\footnote{The isospin is 1/2 for nucleon, but could be defined as 0 for strangeon.}
It seems natural if Nature loves a principle of
flavour maximization~\citep{xu18,2022IJMPE..3150037M}.
This difference carries profound implications: {\bf 2}-flavoured strong matter in bulk is too less-massive to be stable against decaying into atomic nuclei (the minimum mass of conventional neutron star could be $\sim 0.1M_\odot$), while the baryon number of stable strangeon matter could be as low as $10^{10}$ (so-called strangeon nugget).

2, {\it Mass and Wavepacket Size}.
It might be fine to represent an atomic nucleus as a liquid drop since the quantum wavepackets of identical nucleons could overlap.
However, the quantum wavelength of massive strangeon is comparably short, and one may attribute the properties of strangeon matter to classical physics.
Unless during its formation at high temperature, strangeon matter could be in a solid state as the thermal kenematic energy would be much lower than the interaction potential (a few $10$ MeV) between strangeons.
The free energy, elastic and gravitational, of solid strangeon stars could power extraordinary astrophysical bursts, from so-called magnetars~\citep{2006MNRAS.373L..85X}, to fast radio bursts~\citep{2022SCPMA..6589511W}, and even $\gamma$-ray burst~\citep{2009ScChG..52..315X}.
Further, the non-relativity of strangeons results in a stiffer equation of state than that of nucleon matter~\citep{LX2009}, a simple key on solving the hyperon puzzle.

Is strangeon matter just hyperon matter?
No hyperon matter has been conjectured yet, but strangeon differs from hyperon in two facts at least.
$a$, The baryon number of hyperon is $B=1$, whereas strangeon is baryon-like but not a real baryon, whose number could be $B>1$, at least $B=2$ (dibaryon) and maybe $B=6$ (quark alpha, $Q_\alpha$).
$b$, The interaction between two $\Lambda$-particles could be attractive, with binding energy of a few MeVs~\citep{Green2021} or higher according to  lattice QCD simulations. Certainly, an attractive interaction inbetween could not support against gravity.

Let's go back to Fig.~\ref{fig:Cu}.
The mass spectrum of normal electric matter is continuous, from molecules (a few atom-units), to dusts (much smaller than the Avogadro constant), planets ($\sim 10^{52}$), and stars ($\sim 10^{57}$).
However, there could be a gap (around $\sim A_{\rm c}$) in the mass distribution of strong matter: nucleon matter exists for baryon number $A<A_{\rm c}$, while strangeon matter for $A_{\rm c}<A<A_{\rm max}$.

\begin{figure*}
\centering
\includegraphics[scale=0.66]{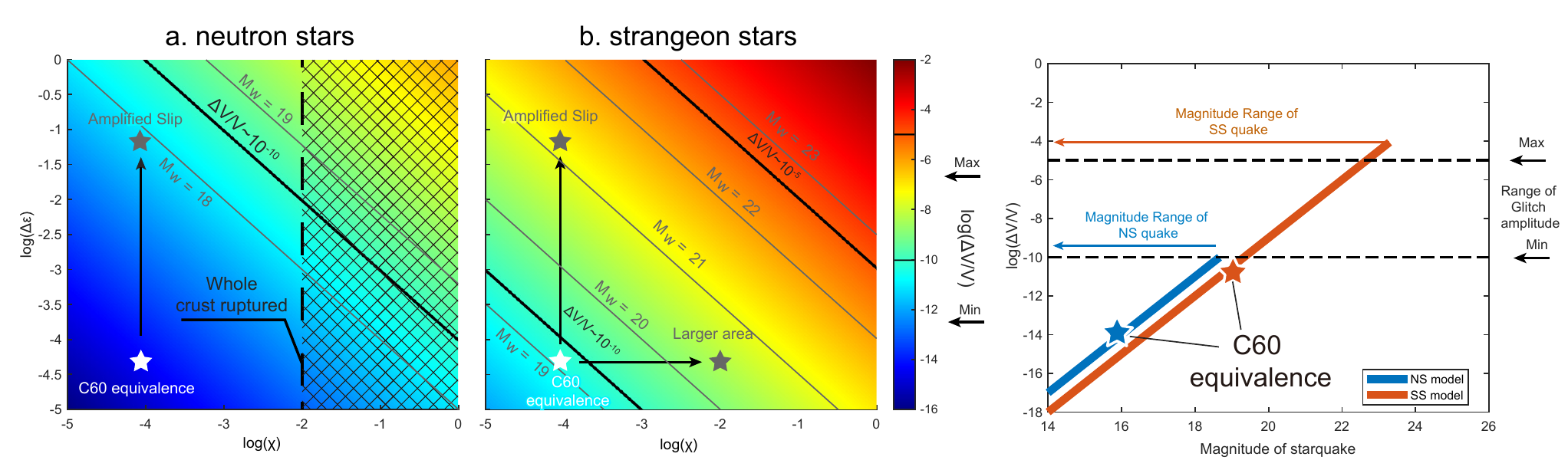}
\caption{A comparison between starquakes in conventional neutron stars (panel ``a'', NS) and in strangeon stars (panel ``b'', SS), showing dependence of glitch magnitude (log ($\Delta\nu/\nu$)) on the starquake strain drop ($\varepsilon$) and moment ratio ($\chi$).
The maximum and minimum glitch magnitude ($10^{-10}\sim10^{-5}$) are bounded by black lines. The line of $10^{-5}$ is not given in NS model for it is over the theoretical maximum.
The shadow in panel ``a'' mark the area that $\chi$ is not allowed in neutron stars. A scaled earthquake of the largest observed one (i.e. C60: 1960 Chile earthquake) is marked as white stars in both figures as references, respectively.
Grey lines are isolines of starquake magnitude with corresponding $\Delta\varepsilon$ and $\chi$.
Panel ``c'' shows the comparison between NS and SS models. The range of observation glitch amplitude is shown as horizontal dot lines. C60 in the two models is respectively marked as the blue and orange star.
These calculations presented by~\cite{LuRP2022} show that the observed glitch amplitude can be explained by the starquakes in the strangeon star model, though the required scaled
starquake magnitude is much larger than that occurred on the Earth.}
\label{quake}
\end{figure*}

Quakes are natural on the solid Earth, and thus on solid strangeon stars too.
Similar to electric matter inside the solid Earth, the geometrical symmetry of strangeons in solid matter could also not be simple and perfect, and a quasicrystal structure is possible too.
An amorphous (glass-like) structure of solid strangeon matter could even form if the cooling is so quick that phase equilibrium can hardly be achieved during liquid-solid transition.
Anyway, a quake as a result of slip should occur along the direction on a fault plane~\citep{LuRP2022}, as commonly observed on the Earth, with simulated results illustrated in Fig.~\ref{quake}.
Besides strangeon stars, strangeon nuggets formed through a crossover QCD phase transition in the early Universe before the big bang nucleosynthesis has also been demons, indicating  a strangeon dark matter candidate without introducing particles
beyond the standard model~\citep{Wu2022}.

\section{Summary and Outlooks}

This note is my second one in memory of Landau's original idea on neutron star, with the first one published around ten years ago~\citep{XuRX2011}.
Following Landau's approach more than 90 years ago, we are explaining a novel kind of matter at pressure free, the strong matter by strangeons rather than by nucleon, as an analogy of the normal atom matter illustrated in Fig.~\ref{fig:Cu} for an example.
Today, the nature of pulsar-like stars is focused in both physics and astrophysics, hopefully to be the first big problem solved in this multi-messenger era of astronomy.
A detailed study in microphysics may eventually reveal the critical number, $A_{\rm c}$.
Although the number can't yet be computed from the first principles at present, it does not mean that Nature will not adopt this idea and refuse to use it.

In fact, a strangeon is a kind of multi-quark states, which is hotly discussed in high-energy physics, both theoretical and in experimental.
It is surely welcome to investigate the formation of strangeon with either lattice QCD or effective field theory, that would be conducive to quantitatively describing the state equation of supranuclear matter.
Although one cannot rule out other possibilities, e.g., that multi-quark states and free quarks could coexist~\citep{Burikham2009}, the energy scale of strong matter at pressure free should not be fine-tuned.
Therefore, a consistent state of chemical composition would be reasonable, but the baryon numbers of strangeons from the surface to the center may increase slightly nevertheless.

Looking back at the journey to understand the nature of our material world, we could be trying to know strong matter in an opposite direction to that of electric matter.
The study of electric matter in Fig.\ref{fig:Cu}told us the micro-units and their quantum nature, but we are reproducing the world of strong matter with the standard model (QCD and the electroweak theory).

\section*{ACKNOWLEDGMENTS}
The author would like to thank those involved in the continuous discussions in the pulsar group at Peking University. This work is supported by the National SKA Program of China (2020SKA0120100).
The author would like to acknowledge his debt to Mr. Shichuan Chen, who reads the manuscript carefully with valuable comments.


\end{document}